\documentclass[10pt]{article}

\usepackage{amsmath}
\usepackage{amssymb}

\usepackage{graphicx}

\usepackage{cite}

\usepackage{color} 

\usepackage[labelfont=bf,labelsep=period,justification=raggedright]{caption}

\makeatletter
\renewcommand{\@biblabel}[1]{\quad#1.}
\makeatother

\date{}

\begin{document}

\begin{flushleft}
{\Large
\textbf{Quantifying the role of population subdivision in evolution on rugged fitness landscapes}
}
\\
Anne-Florence Bitbol$^{1,\ast}$, 
David J. Schwab$^{1}$
\\
\bf{1} Lewis-Sigler Institute for Integrative Genomics and Department of Physics, Princeton University, Princeton, NJ 08544, USA
\\
$\ast$ E-mail: afbitbol@princeton.edu
\end{flushleft}

\section*{Abstract}
Natural selection drives populations towards higher fitness, but crossing fitness valleys or plateaus may facilitate progress up a rugged fitness landscape involving epistasis. We investigate quantitatively the effect of subdividing an asexual population on the time it takes to cross a fitness valley or plateau. We focus on a generic and minimal model that includes only population subdivision into equivalent demes connected by global migration, and does not require significant size changes of the demes, environmental heterogeneity or specific geographic structure. We determine the optimal speedup of valley or plateau crossing that can be gained by subdivision, if the process is driven by the deme that crosses fastest. We show that isolated demes have to be in the sequential fixation regime for subdivision to significantly accelerate crossing. Using Markov chain theory, we obtain analytical expressions for the conditions under which optimal speedup is achieved: valley or plateau crossing by the subdivided population is then as fast as that of its fastest deme. We verify our analytical predictions through stochastic simulations. We demonstrate that subdivision can substantially accelerate the crossing of fitness valleys and plateaus in a wide range of parameters extending beyond the optimal window. We study the effect of varying the degree of subdivision of a population, and investigate the trade-off between the magnitude of the optimal speedup and the width of the parameter range over which it occurs. Our results, obtained for fitness valleys and plateaus, also hold for weakly beneficial intermediate mutations. Finally, we extend our work to the case of a population connected by migration to one or several smaller islands. Our results demonstrate that subdivision with migration alone can significantly accelerate the crossing of fitness valleys and plateaus, and shed light onto the quantitative conditions necessary for this to occur. 

\section*{Author Summary}
Experimental evidence has recently been accumulating to suggest that fitness landscape ruggedness is common in a variety of organisms. Rugged landscapes arise from interactions between genetic variants, called epistasis, which can lead to fitness valleys or plateaus. The time needed to cross such fitness valleys or plateaus exhibits a rich dependence on population size, since stochastic effects have higher importance in small populations, increasing the probability of fixation of neutral or deleterious mutants. This may lead to an advantage of population subdivision, a possibility which has been strongly debated for nearly one hundred years. In this work, we quantitatively determine when, and to what extent, population subdivision accelerates valley and plateau crossing. Using the simple model of an asexual population subdivided into identical demes connected by gobal migration, we derive the conditions under which crossing by a subdivided population is driven by its fastest deme, thus giving rise to the maximal speedup. Our analytical predictions are verified using stochastic simulations. We investigate the effect of varying the degree of subdivision of a population. We generalize our results to weakly beneficial intermediates and to different population structures. We discuss the magnitude and robustness of the effect for realistic parameter values.

\section*{Introduction}
Natural selection drives populations towards higher fitness (i.e. reproductive success), but crossing fitness valleys or plateaus may facilitate progress up a rugged fitness landscape. Rugged fitness landscapes arise from epistasis, i.e. interactions between genetic variants. For instance, two mutations together can yield a benefit while each of them alone is detrimental: such reciprocal sign epistasis can give rise to a fitness valley~\cite{Dawid10,Draghi13}. While the high dimensionality of genotype space makes it challenging to probe the structure of fitness landscapes~\cite{Franke11,Szendro13}, evidence has been accumulating for frequent landscape ruggedness, especially in recent years~\cite{Whitlock95,Schrag97,Beerenwinkel07,Trindade09,Andersson10,Dawid10,Bloom10,Kryazhimskiy11b,Breen12,Szendro13,Draghi13,Gong13,Covert13,Ostman14}. 

Population structure can play an important role in evolution~\cite{Korona94,Hallatschek07,Waclaw10,Martens11,Martens11b,Otwinowski11,Zhang11,Greulich12,Hermsen12}. In particular, the time taken to cross a fitness valley or plateau depends on population size since stochastic effects such as genetic drift have an increased importance in small populations, allowing neutral and deleterious mutations to fix with increased probability~\cite{Ewens79,Weinreich05,Rozen08,Weissman09}. Population subdivision into demes can allow the maintenance of larger genetic diversity due to increased genetic drift as well as to the quasi-independent explorations of the fitness landscape that are run in parallel by each deme. Subdivision may thereby facilitate valley or plateau crossing locally and subsequent migration can then spread beneficial mutations throughout the entire subdivided population (``metapopulation''). This idea was first discussed by Wright in his shifting balance theory~\cite{Wright31,Wright32,Wright40,Wright82} and the importance of this effect has been the subject of a long debate~\cite{Lande85,Slatkin89,Wade91,Barton93,Coyne97,Gavrilets97,Wade98,Coyne00,Crow08,Wade13}. In this work, we investigate the role of subdivision with global migration alone, without additional effects such as strong dependence of deme size on fitness, including extinction and refounding of demes, which played a crucial role in Wright's theory. Our generic and minimal model enables us to quantatively determine the conditions under which population subdivision accelerates fitness valley or plateau crossing.

Studying quantitatively the effect of subdivision on evolution may help in inferring fitness landscape structure from evolution experiments~\cite{Desai13}. Work on structured populations has been used as qualitative proof of landscape ruggedness~\cite{Korona94}. Current experiments investigating the evolution of subdivided populations at various migration rates have produced mixed results, some demonstrating faster adaptation of subdivided populations~\cite{Kerr, KerrPreprint}, and others not~\cite{Kryazhimskiy11}. It is therefore important to determine under what conditions subdivision accelerates fitness valley or plateau crossing. Additionally, population subdivision is extremely common in natural systems. For instance, evidence has recently been found for compartmentalization of HIV in different organs of a single patient~\cite{vanMarle07,Schnell10}. 

Here we show that subdivision can significantly accelerate fitness valley or plateau crossing over a wide parameter range, both with respect to a non-subdivided population and with respect to a single deme. Intuitively, deleterious or neutral intermediate mutations may fix in individual demes, allowing for the maintenance of a larger proportion of these mutants in a metapopulation than in a well-mixed population. We first determine the optimal speedup of valley or plateau crossing by subdivision, in the best possible scenario where valley or plateau crossing by the metapopulation is driven by that of its fastest deme. This enables us to demonstrate that isolated demes must be in the sequential fixation regime for subdivision to significantly accelerate crossing. We then determine the conditions under which the best possible scenario can be realized. Using Markov chain theory, we obtain analytical expressions for the parameter range where valley or plateau crossing by a metapopulation is as fast as that of its fastest deme. Our analytical predictions are verified using stochastic simulations. Furthermore, we discuss the effect of varying the degree of subdivision of a population, and investigate the trade-off between the magnitude of the optimal speedup and the width of the parameter range over which it occurs. Finally, we extend our work to weakly beneficial mutations and to a population connected to smaller islands, and we discuss the magnitude and robustness of the effect for realistic parameter values.

\section*{Results}

 Our results are organized as follows. First, we specify our model for the evolutionary dynamics of a subdivided population with migration. Then, we focus on the `best possible' scenario where the metapopulation is driven by its fastest deme. We calculate the ratio of the valley-crossing time for the metapopulation to the valley-crossing time for an equally-sized well-mixed population under this strong assumption. This yields the optimal speedup that may be obtained by subdivision, and enables us to demonstrate that sequential fixation in individual demes is necessary to achieve a significant speedup. Then, we determine the range of parameter values for which the best possible scenario is attained, i.e. the valley-crossing time for the metapopulation is indeed dominated by the valley-crossing time of its fastest deme. Qualitatively, migration has to be both rare enough to enable demes to cross the fitness valley or plateau quasi-independently and frequent enough to allow fast spreading of the final beneficial mutation to the whole metapopulation once it has fixed in the fastest deme: these conditions yield an optimal window of migration rates. Finally, we compare our analytical predictions with results from stochastic simulations.

\subsection*{Model of evolutionary dynamics in a subdivided population}

We focus on asexual individuals, characterized by their genotype and associated fitness $f$. Each individual has a division rate proportional to $f$, and a death rate $d$, which is the same for all. We consider an ensemble of $ D $ identical demes, each with a constant number $N$ of individuals. The division rate averaged over the individuals of a deme is thus equal to the death rate $d$. We treat migration as a random exchange of two individuals between two different demes, occurring at rate $2m$ per individual. In our model, exchange between any two demes is equally likely, as in Wright's ``island model''~\cite{Wright31}. This constitutes a generic and minimal model of subdivision with migration, without any dependence of migration rate on the average fitness of a deme (in contrast with models where demes containing beneficial mutants increase significantly in size and migrate more rapidly~\cite{Wright32,Lande85}), or additional effects of extinction and re-founding of demes~\cite{Wright32,Wright82,Lande85}, specific geographic structure~\cite{Korona94,Hallatschek07,Martens11,Martens11b,Otwinowski11}, or spatially heterogeneous environments~\cite{Waclaw10,Zhang11,Greulich12,Hermsen12}, on which previous studies focused. 

We consider the simplest fitness valley or plateau, involving three successive genotypes denoted by `0', `1' and `2' (see Fig.~\ref{Opti}A). The initial genotype is taken as reference for fitness: $f_0= 1$. We denote the fitnesses of the subsequent genotypes by $f_1=1-\delta$ and $f_2=1+s$. The first mutation is assumed to be either neutral ($\delta=0$), which yields a fitness plateau, or deleterious ($\delta>0$), which corresponds to a fitness valley, while the second mutation is assumed to be beneficial ($s>0$). We focus on first mutations that are not too strongly deleterious: $\delta\ll 1$. We only allow forward mutations, and note that including back mutations does not qualitatively affect crossing times~\cite{Weissman09}. Finally, we assume that all mutations have probability $\mu$ per division, but generalization to different mutation probabilities is straightforward. 

In this paper, we focus on the average time $\tau_m$ required for the whole metapopulation to cross the fitness valley or plateau, i.e. to fix mutation `2' in all demes, starting from an initial state where all individuals have genotype `0'.

\subsection*{The best possible scenario}

For small enough migration rates, each deme in the metapopulation performs a quasi-independent trial at crossing the valley or plateau. At best, the valley or plateau crossing time $\tau_m$ of the whole metapopulation is dominated by that, $\tau_c$, of the ``champion'' deme in the metapopulation, i.e. the deme that crosses the fitness valley or plateau fastest.

We now focus on this best possible scenario, which is illustrated schematically in Fig.~\ref{Opti}B: first, the champion deme crosses the valley or plateau by sequential fixation, and then the beneficial mutation rapidly spreads by migration of through the whole metapopulation. Once this best possible scenario is characterized, the crucial question will be whether, and under what conditions, it can be attained: this point will be addressed in the following section.  

\subsubsection*{Determination of $\tau_c$} 

Valley or plateau crossing by a non-structured, well-mixed population can occur by two different mechanisms: sequential fixation and tunneling. The former corresponds to fixation of mutation `1' in the whole population, and to subsequent fixation of the beneficial mutation `2'. Conversely, the latter occurs when the beneficial mutation arises in a small fluctuating minority of first-mutants, and fixes directly: tunneling does not involve fixation of the intermediate mutation `1'~\cite{Weissman09}. For given values of the parameters $\delta$, $s$, and $\mu$, sequential fixation is the fastest process for small populations, where genetic drift plays an important part. Tunneling becomes the dominant process of valley or plateau crossing when the number $N$ of individuals per deme exceeds a threshold value $N_\times$, which depends on $\delta$, $s$, and $\mu$ (see Ref.~\cite{Weissman09} for a full discussion of this threshold value). Fig.~\ref{Opti}C shows simulation results for the valley crossing time $\tau$ of a non-subdivided population versus its size, and illustrates these two different regimes and the transition between them. Note that in our simulations (described in Methods, Sec.~\ref{AppSim}), we hold fixed the carrying capacity $K$ of populations (or demes) instead of the number of individuals $N$. This softer constraint is more realistic and avoids some possible biases in the metapopulation case (see Methods, Sec.~\ref{AppSimK}). In practice, each individual divides at a rate $f(1-N/K)$ and dies at a constant rate $d$: hence, at steady-state, $N \approx K(1-d/f)$. We choose $d=0.1$, and fitnesses $f$ of order one, thus $N\approx0.9K$. 

We now consider $D$ independent demes with no migration, and we determine the crossing time $\tau_c$ of the fastest of these $D$ demes, both for demes in the sequential fixation regime and for demes in the tunneling regime.  

\paragraph*{Demes in the sequential fixation regime.}Let 
\begin{equation}
p_{ij}=\frac{1-e^{f_i-f_j}}{1-e^{N(f_i-f_j)}}\,
\label{pij}
\end{equation}
denote the probability of fixation of genotype `$j$', with fitness $f_j$, starting from a single individual with genotype `$j$' in a deme where all other individuals initially have genotype `$i$' and fitness $f_i\neq f_j$~\cite{Ewens79,Weissman09}. If $f_i=f_j$, the probability of fixation of genotype `$j$' reads $p_{ij}=1/N$. Valley or plateau crossing by sequential fixation involves two successive steps. The first step, fixation of the intermediate mutation `1', occurs with rate $r_{01}=N \mu d p_{01}$, where $N\mu d$ is the total mutation rate in the deme. (Recall that the deme size $N$ is fixed, and that $d$ represents the birth/death rate. Note that the correspondence with Ref.~\cite{Weissman09} is obtained by multiplying by $1/d$ all the timescales in this reference, which are expressed in numbers of generations.) Similarly, the second step, fixation of the final beneficial mutation `2', has rate $r_{12}=N \mu d p_{12}$. The first step is longer than the second one since mutation `1' is neutral or deleterious, while mutation `2' is beneficial. If the first step dominates, the distribution of crossing times is approximately exponential with rate $r_{01}$. The shortest crossing time among $ D $ independent demes is then distributed exponentially with rate $ D r_{01}$ (see Methods, Sec.~\ref{AppB}). Thus, the average crossing time of the champion deme reads $\tau_c\approx(D r_{01})^{-1}$. Denoting by $\tau_{id}\approx r_{01}^{-1}$ the average crossing time for an isolated deme, we obtain
\begin{equation}
\frac{\tau_c}{\tau_{id}}\approx\frac{1}{D}\,.
\label{Dfac}
\end{equation}
Hence, the champion deme crosses the valley $ D $ times faster on average than a single deme. This simple result holds for $ D p_{01}\ll p_{12}$. For simplicity, we restrict ourselves to this regime in the main text, but we provide the general method for calculating $\tau_c$ in Methods, Sec.~\ref{AppB}. We use this general method to calculate numerically the exact value of $\tau_c$ in our examples below.

\paragraph*{Demes in the tunneling regime.}Assuming that $N \mu<1$, so that there is no competition between different mutant lineages, valley or plateau crossing by tunneling involves a single event with constant rate, namely the appearance of a ``successful'' `1'-mutant, whose lineage includes a `2'-mutant that fixes~\cite{Weissman09}. Crossing time is thus exponentially distributed. Therefore, in this case too, the crossing time $\tau_c$ of the champion deme among $D$ isolated demes is $D$ times smaller than that of an average isolated deme (see Methods, Sec.~\ref{AppB}): Eq.~\ref{Dfac} is valid in the tunneling regime too.

\subsubsection*{Sequential fixation in individual demes is necessary for significant speedups}

\begin{figure}[h t b]
\centering
\includegraphics[width=\textwidth]{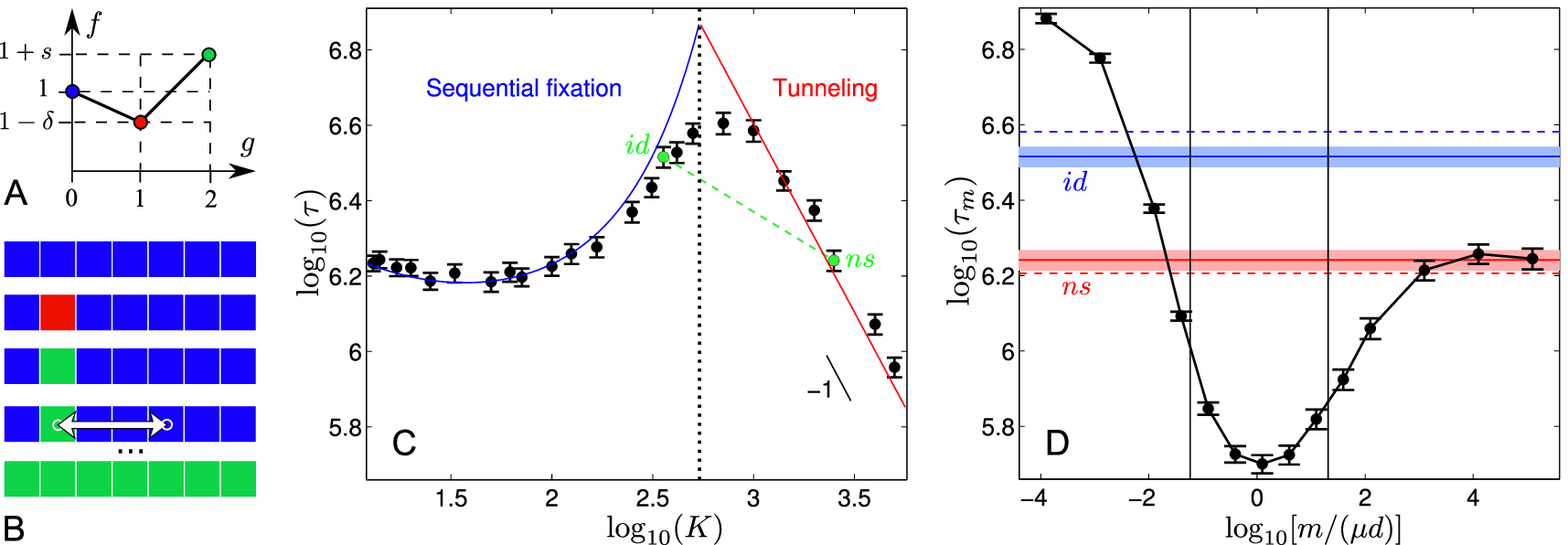}
\caption[Population subdivision with migration can accelerate fitness valley crossing.]{\label{Opti} \textbf{Population subdivision with migration can accelerate fitness valley crossing.} 
\textbf{A.} Fitness valley: fitness $f$ versus genotype $g$. \textbf{B.} Schematic representation of the best possible scenario, for a metapopulation with $D=7$ demes. Each square represents a deme of identical size, and a row represents the metapopulation. Colors represent genotypes, with the color-code defined in A. Initially (top row), all demes have genotype `0'. The demes explore the fitness landscape described in A quasi-independently, and one of them, the ``champion'' deme (second from the left here), crosses the fitness valley first (second and third row). Individual demes are assumed to be in the sequential fixation regime, so this deme fixes first mutation `1' and then mutation `2'. The beneficial mutation `2' then spreads by migration, which is modeled by random exchange of individuals between demes (arrow on the fourth row), leading to fixation of mutation `2' in the whole metapopulation (fifth row). \textbf{C.} Average valley crossing time $\tau$ of a non-structured population, as a function of its carrying capacity $K$, in logarithmic scale. Dots are simulation results, averaged over 1000 runs for each value of $K$; error bars represent 95\% confidence intervals (CI). Theoretical predictions from Ref.~\cite{Weissman09} are plotted for the sequential fixation regime (blue line) and for the tunneling regime (red line), using $N= 0.9 K$ (see text) to make the correspondence. The transition between these two regimes is indicated by a dotted line. The carrying capacities at stake in D are highlighted in green ($id$: isolated deme; $ns$: non-subdivided population). Parameter values: $d=0.1$, $\mu=8\times10^{-6}$, $s=0.3$ and $\delta=6\times 10^{-3}$. \textbf{D.} Average valley crossing time $\tau_m$ of a metapopulation composed of $ D =7$ demes each with carrying capacity $K=357$ (total carrying capacity: $DK=2499$), plotted versus the migration-to-mutation rate ratio $m/(\mu d)$, in logarithmic scale. Parameter values are the same as in C, and only the migration rate $m$ is varied. Dots represent simulation results averaged over 1000 runs for each value of $m$, and error bars are 95\% CI. Black vertical lines represent the limits of the interval of $m/(\mu d)$ in Eq.~\ref{interv}. Blue (resp. red) line: valley crossing time for an isolated deme ($id$) with $K=357$ (resp. a non-subdivided population ($ns$) with $K=2500$) for the same parameter values, averaged over 1000 runs; shaded regions: 95\% CI. Dashed blue (resp. red) lines: corresponding theoretical predictions from Ref.~\cite{Weissman09} (see C).}
\end{figure}

In the best possible scenario, where crossing by the metapopulation is dominated by that of the champion deme, i.e. $\tau_m\approx\tau_c$, the previous paragraph shows that $\tau_m/\tau_{id}\approx1/ D $, both when isolated demes are in the sequential fixation regime and when they are in the tunneling regime. Hence, it is necessary to have 
\begin{equation}
\frac{\tau_{id}}{\tau_{ns}}<D\,,\label{cn}
\end{equation}
where $\tau_{ns}$ is the average crossing time of the non-subdivided population, for subdivision to speed up valley or plateau crossing in the best scenario (i.e. for $\tau_m\approx\tau_c$ to be smaller than $\tau_{ns}$).  This necessary condition is general since it holds \emph{a fortiori} beyond the best scenario. Graphically, in Fig.~\ref{Opti}C, which is a logarithmic plot of crossing time versus population size for a non-structured population, the slope of the line joining the isolated deme to the non-subdivided population has to be less negative than -1 in order for speedups to be possible. Recall indeed that the nonsubdivided population is $D$ times larger than an isolated deme. The necessary condition in Eq.~\ref{cn} leaves the possibility of significant speedups in the non-trivial case where a single isolated deme crosses slower than a non-subdivided population ($\tau_{id}>\tau_{ns}$). Fig.~\ref{Opti}D demonstrates a significant speedup by subdivision obtained in this regime where $1<\tau_{id}/\tau_{ns}<D$.  

Let us consider a metapopulation such that isolated demes are in the tunneling regime. Then, the larger non-subdivided population with $ND$ individuals is also in the tunneling regime~\cite{Weissman09}. Assuming that $N D\mu<1$, valley or plateau crossing by this non-subdivided population follows the same laws as crossing by the demes. Since the average crossing time by tunneling is inversely proportional to population size (see Ref.~\cite{Weissman09} and Fig.~\ref{Opti}C), we obtain $\tau_{id}/\tau_{ns}=D$, in contradiction with Eq.~\ref{cn}. This implies that, even in the best possible scenario, subdivision cannot accelerate crossing if isolated demes are in the tunneling regime (since here, $\tau_m/\tau_{ns}\approx1$). Thus, having isolated demes in the sequential fixation regime is a necessary condition for subdivision to accelerate crossing. Importantly, however, the non-subdivided population is not required to be in the sequential fixation regime. For instance, in Fig.~\ref{Opti}D, the non-subdivided population is in the tunneling regime. Note that when $N D\mu>1$, the population enters the semi-deterministic regime~\cite{Weissman09} and the average crossing time need not be proportional to $1/N$. Minor speedups may exist in this regime, but such effects are beyond the scope of this work. In all the following, we will focus on the regime $ND \mu<1$.

\subsubsection*{Maximal possible speedup by subdivision}

The speedup gained by subdividing a population of a given total size is directly described by the ratio $\tau_m/\tau_{ns}$ of the valley crossing time of a metapopulation to that of a non-subdivided population. Here, we discuss the values this ratio can take in the best possible scenario, where valley crossing by the metapopulation is dominated by that of the champion deme, and we determine the valley depth for which the highest speedups are obtained (i.e. for which this ratio is smallest). 

Let us first focus on the case where both the non-subdivided population and the isolated deme are in the sequential fixation regime. The average valley crossing time by the champion deme reads $\tau_c\approx1/( D N\mu d p_{01})$ (see our calculation of $\tau_c$ above). In the best possible scenario, $\tau_m\approx\tau_c$. The average valley crossing time by the non-subdivided population is $\tau_{ns}\approx1/( D N\mu d p'_{01})$, where $p'_{01}=(e^{\delta}-1)/(e^{ D N\delta}-1)$ is the fixation probability of an individual with genotype `1' in a population of $N D$ individuals where all the others initially have genotype `0' (see Eq.~\ref{pij}). Hence, we obtain 
\begin{equation}
\frac{\tau_m}{\tau_{ns}}\approx \frac{p'_{01}}{p_{01}}=\frac{e^{N\delta}-1}{e^{ D N\delta}-1}\,. \label{spup}
\end{equation}
In the case of a plateau, this reduces to $\tau_m/\tau_{ns}=1/ D$. These results demonstrate that if both the non-subdivided population and the isolated deme are in the sequential fixation regime, then subdivision significantly accelerates crossing in the best scenario. The speedup by subdivision becomes larger (i.e. $\tau_m/\tau_{ns}$ becomes smaller) when the number of demes $ D$ is increased at fixed valley depth $\delta$ and fixed deme size $N$ (or fixed total population size $\mathcal{N}=N D$). Besides, for $ D\gg 1$, the ratio in Eq.~\ref{spup} decreases when $\delta$ is increased at fixed $N$ and $ D$: the highest speedups are obtained for the deepest valleys. However, as $\delta$ is increased, the non-subdivided population will eventually enter the tunneling regime (see Fig.~\ref{Opti}C). 

Let us now consider the alternative case, where the non-subdivided population is in the tunneling regime, while the isolated demes are in the sequential fixation regime. In this case, $\tau_c=1/(ND \mu d p_{01})$, where $p_{01}$ is the fixation probability of a `1'-mutant in an isolated deme (see Eq.~\ref{pij}), while $\tau_{ns}=1/(ND\mu d q)$, where $q$ is the probability that a `1'-mutant is ``successful'' in the tunneling process, i.e. that its lineage includes a `2'-mutant that fixes in the non-subdivided population~\cite{Weissman09}. Hence, in the best scenario, where $\tau_m\approx\tau_c$, we obtain
\begin{equation}
\frac{\tau_m}{\tau_{ns}}\approx\frac{q}{p_{01}}\,.\label{neweq}
\end{equation}
Since $q$ is independent from population size~\cite{Weissman09}, it also represents the probability of successful tunneling \emph{in an isolated deme}. For isolated demes in the sequential fixation regime, $q<p_{01}$ by definition~\cite{Weissman09}. Hence, Eq.~\ref{neweq} entails $\tau_m/\tau_{ns}<1$. Thus, speedups always exist in the best scenario, provided that the necessary condition that isolated demes cross the plateau by sequential fixation is satisfied. In the case of a fitness plateau, $q=\sqrt{\mu s}$~\cite{Weissman09}, while $p_{01}=1/N$. Hence, Eq.~\ref{neweq} yields
\begin{equation}
\frac{\tau_m}{\tau_{ns}}\approx N \sqrt{\mu s} \,.\label{lemin3}
\end{equation}
In the other extreme case of a sufficiently deep valley that satisfies $\delta\gg 2\sqrt{\mu s}$, we have $q=\mu s/\delta$~\cite{Weissman09}. Using the condition $\delta\ll1$, Eq.~\ref{pij} yields $p_{01}=\delta/(e^{N\delta}-1)$. Hence, Eq.~\ref{neweq} gives
\begin{equation}
\frac{\tau_m}{\tau_{ns}}=\mu s\frac{e^{N\delta}-1}{\delta^2}\,. \label{unautrelabel}
\end{equation}
Interestingly, these expressions of $\tau_m/\tau_{ns}$ are independent of $ D$ at fixed $N$. This stands into contrast with the regime discussed above where the non-subdivided population is in the sequential fixation regime. At fixed $N$, the ratio $\tau_m/\tau_{ns}$ expressed in Eq.~\ref{unautrelabel} is minimal for 
\begin{equation}
\delta\approx \frac{1.594}{N}\,.\label{lemin}
\end{equation}
The minimum of $\tau_m/\tau_{ns}$, corresponding to the largest speedup by subdivision, is obtained for this value of $\delta$: 
\begin{equation}
\frac{\tau_m}{\tau_{ns}}\approx 1.544 \,N^2\mu s \,.\label{lemin2}
\end{equation}
The small values of mutation probabilities in nature ensure that the values of $\tau_m/\tau_{ns}$ in Eqs.~\ref{lemin3} and~\ref{lemin2} can be very small.

\subsection*{Conditions for subdivision to maximally accelerate valley or plateau crossing}

The previous section was dedicated to the study of the best possible scenario, where the valley or plateau crossing time $\tau_m$ of the whole metapopulation is dominated by that, $\tau_c$, of the champion deme in the metapopulation (i.e. the one that crosses fastest). We now determine analytically the conditions under which this best possible scenario is attained. For this, we focus on migration rates much smaller than division/death rates, $2m\ll d$, such that fixation or extinction of a mutant lineage in a deme is not perturbed by migration. In addition, we assume that isolated demes are in the sequential fixation regime, since we showed above that it is a necessary condition for subdivision to significantly accelerate crossing, and that it is a sufficient condition for subdivision to accelerate crossing in the best scenario. 

In a nutshell, migration must be rare enough for demes to evolve quasi-independently, but frequent enough to spread the beneficial mutation rapidly. The analytical results below allow for predicting the range of migration rates such that subdivision maximally accelerates valley or plateau crossing.

\subsubsection*{First condition: quasi-independence} 
Migration must be rare enough for demes to remain shielded from migration while they harbor the intermediate mutation. Hence, the average time for a deme of `1'-mutants to fix the beneficial mutation `2', which reads $\tau_{12}=1/r_{12}=1/(N \mu d p_{12})$, must be smaller than the average extinction time, $t_{e}$, for a deme of `1'-mutants to be wiped out by migration from other demes with genotype `0'. The total rate of migration events in the metapopulation is $ D  N m$, so $t_{e}=n_{e}/( D  N m)$, where $n_e$ is the average total number of migration events required for the `1'-mutants to go extinct. The first condition, $\tau_{12}< t_e$, thus yields
\begin{equation}
\frac{m}{\mu d}<\frac{n_e p_{12}}{ D }\,.
\label{TS}
\end{equation}

Let us now estimate $n_e$. If one deme has fixed genotype `1' while all the others still have genotype `0', the probability that a migration event involves the mutant deme is $p_r=2/ D$. Following such a ``relevant" migration event, extinction of the mutant (`1') lineage occurs if the `0' migrant fixes in the `1' deme while the `1' migrant does not fix in the `0' deme: this occurs with probability $p_{10}(1-p_{01})$. Conversely, the number of mutant demes increases to two with probability $p_{01}(1-p_{10})$, and otherwise remains constant. For $N\delta\gg1$, using also $\delta\ll1$, we have $p_{01}\approx\delta e^{-N\delta}\ll p_{10}\approx \delta$ (see Eq.~\ref{pij}). Hence, migration-induced increases in the number of mutant demes can be neglected, and we obtain
\begin{equation}
n_e\approx\frac{1}{p_rp_{10}(1-p_{01})}\approx \frac{ D}{2\delta}\,.
\label{nd}
\end{equation}

In Methods, Sec.~\ref{AppA}, we derive the general expression of $n_{e}$, which does not require $N\delta\gg 1$, using finite Markov chain theory~\cite{Ewens79}. Note that this general expression is important because subdivision generically most accelerates valley crossing for $N\delta\approx1$ (see Eq.~\ref{lemin}).

\subsubsection*{Second condition: rapid spreading} 
Migration must be frequent enough for the average spreading time $t_s$ of the final mutation through the whole metapopulation to be shorter than the valley or plateau crossing time $\tau_c\approx1/( D r_{01})$ by the champion deme. Let $n_s$ be the average number of migration events required for the final beneficial mutants (with genotype `2') to spread by migration, once the champion deme has fixed genotype `2'. Then, we can write $t_s=n_{s}/( D  N m)$, and the second condition reads
\begin{equation}
n_s p_{01}<\frac{m}{\mu d}\,.
\label{TI}
\end{equation}

Let us now estimate $n_s$, starting from a state where the champion deme has fixed genotype `2', while all others still contain genotype `0'. (Note that some demes may have genotype `1', but this is rare since fixation of mutation `1' is the slowest step. Moreover, this would not change the spreading time for a plateau and would shorten it for a valley.) Let us focus on the regime where $s\ll1$ but $Ns\gg1$, such that mutation `2' is substantially, but not overwhelmingly, beneficial~\cite{Weissman09}. As in the above discussion about $n_e$, we then obtain $p_{20}\ll p_{02}$. Thus, it is possible to neglect any migration-induced decrease in the number of demes with genotype `2', which we denote by $i$. The probability that a migration step exchanges individuals with different genotypes is $p_i=2i( D-i)/[ D( D-1)]$, and the probability that such a relevant migration step increases $i$ by one is $p_{02}\approx s$. Hence, we obtain
\begin{equation}
n_s\approx\sum_{i=1}^{ D-1}\frac{1}{p_i s}=\frac{ D-1}{s}\sum_{i=1}^{ D-1}\frac{1}{i}\approx \frac{ D \log  D}{s}\,,
\label{ns}
\end{equation}
where the last expression is obtained for $ D\gg 1$. In Methods, Sec.~\ref{AppA}, we use finite Markov chain theory to derive the general analytical expression for $n_{s}$, which does not require $Ns\gg 1$. 

\subsubsection*{Combination of the two conditions} 
Together, Eqs.~\ref{TS} and~\ref{TI} yield the interval of $m/\mu d$ over which subdivision maximally accelerates valley or plateau crossing. For
\begin{equation}
 n_s p_{01}\ll\frac{m}{\mu d}\ll\frac{n_e p_{12}}{ D }\,,
\label{interv}
\end{equation}
we expect the valley or plateau crossing time $\tau_m$ of the whole metapopulation to be dominated by that of the champion deme: $\tau_m/\tau_{id}\approx1/ D $, where $\tau_{id}\approx r_{01}^{-1}$ is the average crossing time for an isolated deme, and in the best scenario, $\tau_m\approx\tau_c\approx\tau_{id}/ D$. 

In the regime where $Ns, N\delta\gg 1$ and $s,\delta\ll1$, we can use the simple expressions of $n_e$ and $n_s$ given in Eqs.~\ref{nd} and \ref{ns}, which yields
\begin{equation}
\frac{\delta e^{-N\delta}}{s}  D \log  D\ll\frac{m}{\mu d}\ll\frac{1}{2}\left(1+\frac{s}{\delta}\right)\,.
\label{intervs}
\end{equation}
The ratio, $R$, of the upper to lower bound in Eq.~\ref{intervs} reads
\begin{equation}
R=\frac{1}{2D\log D}\,\frac{s}{\delta}\left(1+\frac{s}{\delta}\right) e^{N\delta}\,.\label{unlabela}
\end{equation}
This ratio increases exponentially with $N$ (this dependence on $N$ comes from that of $p_{01}$). This entails that, in this regime, the interval of $m/(\mu d)$ where subdivision most accelerates crossing becomes wider as $N$ increases. However, the width of this interval is limited by the fact that isolated demes have to be in the sequential fixation regime (see Discussion). While the expressions of the interval bounds in Eq.~\ref{intervs} are more illuminating and easier to derive than the general ones, the latter, given in Methods, Sec.~\ref{AppA}, actually play important roles since the highest speedups of valley crossing gained by subdivision are generically obtained for $N\delta\approx 1$ (see Eq.~\ref{lemin}). 

\subsubsection*{Case of the fitness plateau}

We have obtained an explicit expression of the interval of $m/\mu d$ over which subdivision maximally accelerates valley crossing in the case of a relatively deep fitness valley where $N\delta \gg 1$ while $\delta\ll 1$. In the opposite limit of a fitness plateau ($\delta=0$), retaining the assumptions $N s\gg 1$ and $s\ll 1$, Eq.~\ref{interv} can also be simplified. For this, we use the expression of $n_e$ obtained in Eq.~\ref{n1n0} of Methods, Sec.~\ref{AppA}, and the expression of $n_s$ in Eq.~\ref{ns}, and we note that, since mutation `1' is neutral, $p_{01}=1/N$ and $p_{12}=p_{02}\approx s$. Eq.~\ref{interv} then becomes:
\begin{equation}
\frac{1 }{Ns} D\log  D\ll\frac{m}{\mu d}\ll\frac{N s }{2}\log  D\,,
\label{intervsp}
\end{equation}
where we have used $N\gg 1$ and $ D  \gg 1$. The ratio, $R$, of the upper to lower bound in Eq.~\ref{intervsp} reads
\begin{equation}
R=\frac{N^2 s^2}{2 D}\,.\label{unlabel}
\end{equation}
This simple expression of $R$ demonstrates that the range of $m/(\mu d)$ over which subdivision maximally accelerates plateau crossing increases as the deme size $N$ becomes larger, and that this range is quite wide as long as the number of demes satisfies $ D\ll (Ns)^2$, which is a realistic condition (recall that we are in the regime $Ns\gg1$).

\subsection*{Simulation results} 
We now present numerical simulations of the evolutionary dynamics described above, which enable us to test our analytical predictions, and to gain additional insight in the process beyond the optimal scenario. Our simulations are based on a Gillespie algorithm~\cite{Gillespie76,Gillespie77}, and described in detail in Methods, Sec.~\ref{AppSim}. 

Let us first focus on the example presented in Fig.~\ref{Opti}D, which shows an example plot of $\tau_m$ as a function of the ratio of migration to mutation rates, $m/(\mu d)$, obtained through our simulations when varying only the migration rate. With the parameter values used in this figure, the interval of Eq.~\ref{interv} is $5.8\times10^{-2}\ll m/(\mu d)\ll 21$. Note that here, and in the following examples, we use the general expressions of $n_s$ and $n_e$ given in Methods, Sec.~\ref{AppA}, to compute the interval of Eq.~\ref{interv}. Fig.~\ref{Opti}D features a minimum right at the center of this theoretically predicted optimal interval. Moreover, this minimum corresponds to $\tau_m=(5.02\pm0.14)\times 10^5$, while $\tau_{id}=(3.28\pm0.10)\times10^6$: hence, the metapopulation crosses the valley on average 6.54 times faster than an isolated deme. This is very close to the limit of the best possible scenario, where the metapopulation would cross 7 times faster than an isolated deme (since $D=7$ here). This example illustrates that speedups tend towards those predicted in the best scenario, when the interval in Eq.~\ref{interv} is sufficiently wide (here the ratio between its upper and its lower bound is 359). Besides, $\tau_{ns}=(1.74\pm0.05)\times 10^6$ here: comparing it to the above-mentioned value of $\tau_m$ yields a 3.47-fold speedup of valley crossing by subdivision. The simulation results in Fig.~\ref{Opti}D also show that significant (albeit smaller) speedups exist beyond the optimal parameter window.

\begin{figure}[h t b]
\centering
\includegraphics[width=0.5\textwidth]{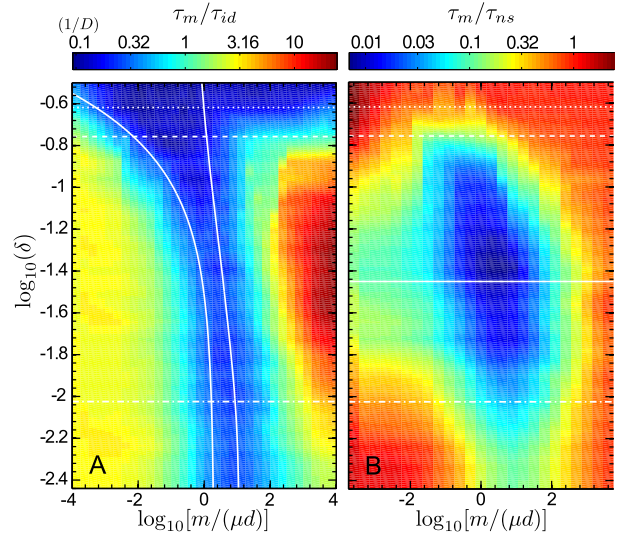}
\caption[Effect of subdivision on valley crossing time as a function of migration rate and valley depth.]{\label{PhaseDiags}\textbf{Effect of subdivision on valley crossing time for various migration rates and valley depths.}\textbf{A.} Heatmap of the ratio $\tau_m/\tau_{id}$ of the average valley crossing time $\tau_m$ of a metapopulation  with $ D =10$ and $K=50$ to that $\tau_{id}$ of an isolated deme with $K=50$, as a function of valley depth $\delta$ and migration-to-mutation rate ratio $m/(\mu d)$, in logarithmic scale. All numerical results are averaged over 100 simulation runs, and the heatmap is interpolated. Solid lines: bounds of the interval in Eq.~\ref{interv}. Dashed line: value of $\delta$ above which a non-subdivided population crosses the valley faster than an isolated deme. Dotted line: value of $\delta$ above which an isolated deme is in the tunneling regime. Dash-dotted line: value of $\delta$ above which the non-subdivided population is in the tunneling regime. Parameter values: $d=0.1$, $\mu=5\times10^{-6}$, $s=0.3$; $\delta$ and $m$ are varied. \textbf{B.} Similar heatmap for the ratio $\tau_m/\tau_{ns}$ of the average valley crossing time $\tau_m$ of a metapopulation to that $\tau_{ns}$ of a non-subdivided population (with $K=500$). Solid line: predicted value of $\delta$, from Eq.~\ref{lemin}, for which the largest speedup by subdivision is expected.}
\end{figure}

Fig.~\ref{PhaseDiags} shows heatmaps of the valley crossing time of a metapopulation as a function of the migration-to-mutation rate ratio, $m/(\mu d)$ (varied by varying $m$), and of the fitness valley depth, $\delta$. Fig.~\ref{PhaseDiags}A shows that the optimal interval of Eq.~\ref{interv} (solid lines) describes well the region where the ratio $\tau_m/\tau_{id}$ of the crossing time of the metapopulation to that of an isolated deme is smallest and tends to the best-scenario limit $1/ D $. For migration rates lower than those in this interval, the ratio $\tau_m/\tau_{id}$ increases when $m$ decreases. This can be understood qualitatively by noting that if $m=0$, $\tau_m$ is determined by the valley crossing time of the slowest among the independent demes. In the opposite case of migration rates larger than those in the optimal interval, $\tau_m$ increases with $m$, and it tends to the non-subdivided case, $\tau_{ns}$, at high values of $m$, as expected. Above a threshold value of $\delta$ (dashed line), $\tau_{ns}$ becomes smaller than $\tau_{id}$, in which case large values of $m$, such that $\tau_m$ tends to $\tau_{ns}$, give a low $\tau_m/\tau_{id}$ (see Fig.~\ref{PhaseDiags}A).

Fig.~\ref{PhaseDiags}B plots the ratio $\tau_m/\tau_{ns}$ of the crossing time of the metapopulation to that of the non-subdivided population, which directly yields the speedup obtained by subdividing a population. It shows that, for the parameter values chosen, subdivision accelerates valley crossing over a large range of valley depths and migration rates, extending far beyond the optimal range given by Eq.~\ref{interv}, and that the metapopulation can cross valleys orders of magnitude faster than a single large population. In addition, above a second, larger threshold value of $\delta$ (dotted line in Fig.~\ref{PhaseDiags}), isolated demes enter the tunneling regime~\cite{Weissman09}: Fig.~\ref{PhaseDiags}B shows that sufficiently above this threshold, the metapopulation no longer crosses the valley faster than the non-subdivided population, as predicted above. While having isolated demes in the sequential fixation regime is a necessary condition to obtain significant speedups by subdivision, the non-subdivided population is not required to be in the sequential fixation regime (see above, and Fig.~\ref{Opti}C-D). The value of $\delta$ above which the non-subdivided population enters the tunneling regime is indicated by a dash-dotted line in Fig.~\ref{PhaseDiags}: significant speedups are obtained both below and above this line. The highest speedups are actually obtained above it, i.e. when the non-subdivided population is in the tunneling regime. With the parameter values used, Eq.~\ref{lemin} predicts a minimum of $\tau_m/\tau_{ns}$ for $\delta\approx 0.035$ (solid line in Fig.~\ref{PhaseDiags}B), which agrees very well with the results of our numerical simulations. (Note that this value of $\delta$ satisfies $\delta\gg 2\sqrt{\mu s}$, and is such that the non-subdivided population is in the tunneling regime. These conditions were used in our derivation of Eq.~\ref{lemin}.)

\section*{Discussion}
\subsection*{Limits on the parameter range where subdivision maximally accelerates crossing} 
In the Results section, we have shown that having isolated demes in the sequential fixation regime is a necessary condition for subdivision to significantly accelerate crossing. This requirement limits the interval of the ratio $m/(\mu d)$ over which the highest speedups by subdivision are obtained. The extent of this interval can be characterized by the ratio, $R$, of the upper to lower bound in Eq.~\ref{interv}. Let us express the bound on $R$ imposed by the requirement of sequential fixation in isolated demes. 

If $2\sqrt{\mu s}\ll\delta\ll 1$, the threshold value $N_\times$ below which an isolated deme is in the sequential fixation regime satisfies $e^{N_\times\delta}\approx\delta^2/(\mu s)$~\cite{Weissman09}. Let us also assume that $N\delta\gg 1$, and that $s\ll 1$ while $Ns\gg 1$, to be in the domain of validity of Eqs.~\ref{intervs} and~\ref{unlabela}. Combining the condition $N<N_\times$ with the expression of $R$ in Eq.~\ref{unlabela} yields
\begin{equation}
R<\frac{\delta}{2\mu \, D \log  D}\left(1+\frac{s}{\delta}\right)\,.
\label{la1}
\end{equation}

For plateaus, isolated demes are in the sequential fixation regime if their size $N$ is smaller than $N_\times=1/\sqrt{\mu s}$~\cite{Weissman09}. In the regime of validity of Eqs.~\ref{intervsp} and~\ref{unlabel} ($s\ll 1$ while $Ns\gg 1$, and $N\gg 1$, $ D\gg 1$), this condition can be combined with Eq.~\ref{unlabel}, which yields
\begin{equation}
R<\frac{s}{2\mu  D}\,. 
\label{la2}
\end{equation}

Both Eq.~\ref{la1} and Eq.~\ref{la2} show that increasing the number $ D$ of demes decreases the range where the highest speedup by subdivision is reached. This is because having more subpopulations makes the spreading of the beneficial mutation slower. In addition, we find that the bound on $R$ is proportional to $1/\mu$. Hence, despite this bound, the interval where subdivision most accelerates plateau crossing can span several orders of magnitude, given the small values of the actual mutation probabilities $\mu$ in nature.

\subsection*{Effect of varying the degree of subdivision of a metapopulation} 

An interesting question raised by our results regards the optimal degree of subdivision. Given a certain total metapopulation size, into how many demes should it be subdivided in order to obtain the highest speedup possible? We first attack this question using our analytical results, and then we present simulation results, which allow for going beyond the best scenario and its associated parameter window.

\begin{figure}[h t b]
\centering
\includegraphics[width=\textwidth]{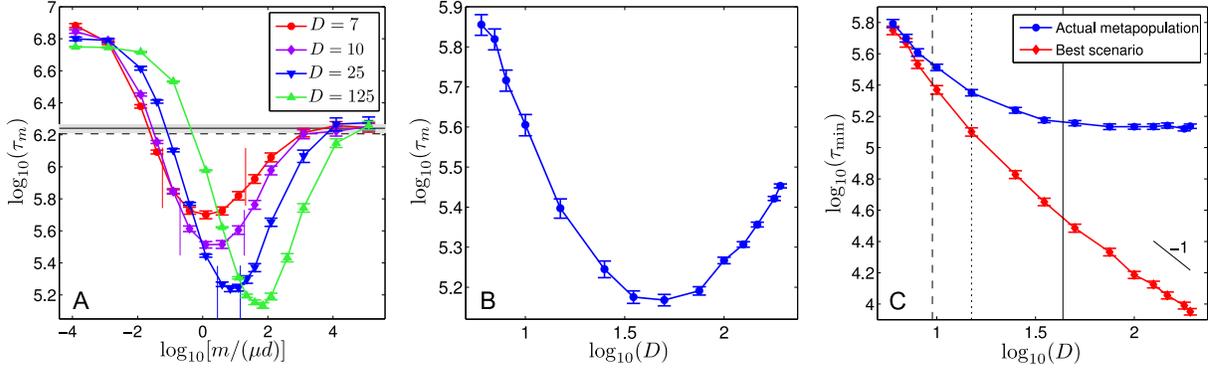}
\caption[Varying the degree of subdivision.]{\label{OptiSub}\textbf{Varying the degree of subdivision of a metapopulation.} 
\textbf{A.} Valley crossing time $\tau_m$ of a metapopulation with total carrying capacity $D K=2500$, versus migration-to-mutation rate ratio $m/(\mu d)$, for four different numbers $D$ of demes. Dots are simulation results, averaged over 1000 runs for each value of $m/(\mu d)$ (500 runs for a few points far from the minima); error bars represent 95\% CI. Vertical lines represent the limits of the interval of $m/(\mu d)$ in Eq.~\ref{interv} in each case, except for $D=125$, where this interval does not exist. Black horizontal line: plateau crossing time for a non-subdivided population with $K=2500$ for the same parameter values, averaged over 1000 runs; shaded regions: 95\% CI. Dashed line: corresponding theoretical prediction from Ref.~\cite{Weissman09}. Parameter values: $d=0.1$, $\mu=8\times10^{-6}$, $s=0.3$ and $\delta=6\times 10^{-3}$ (same as in Fig.~\ref{Opti}C-D); $m$ is varied. \textbf{B.} Valley crossing time $\tau_m$ of a metapopulation with total carrying capacity $D K=2500$, versus the number $D$ of demes, for $m=10^{-5}$ (i.e. $m/(\mu d)=12.5$). Dots are simulation results, averaged over 1000 runs for each value of $D$; error bars represent 95\% CI. Parameter values: same as in A. \textbf{C.} Valley crossing time $\tau_\textrm{min}$, minimized over $m$ for each value of $D$, of a metapopulation with total carrying capacity $D K=2500$, versus the number $D$ of demes. For each value of $D$, the valley crossing time of the metapopulation was computed for several values of $m$, different by factors of $10^{0.25}$ or $10^{0.5}$ in the vicinity of the minimum (see A): $\tau_\textrm{min}$ corresponds to the smallest value obtained in this process. Results obtained for the actual metapopulation (blue) are compared to the best-scenario limit (red) where $\tau_\textrm{min}=\tau_{id}/D$, calculated using the value of $\tau_{id}$ obtained from our simulations. Dots are simulation results, averaged over 1000 runs for each value of $D$; error bars represent 95\% CI. Dashed line: value of $D$ such that $R=100$. Dotted line: value of $D$ above which the deleterious mutation is effectively neutral in the isolated demes. Solid line: value of $D$ such that $R=1$. Parameter values: same as in A and B.}
\end{figure}

Let us consider a metapopulation of given total size $\mathcal{N}=N D$. Our analytical results show that increasing subdivision, i.e. increasing the number $D$ of subpopulations at constant $\mathcal{N}$, leads to stronger speedups of valley crossing (see Eqs.~\ref{spup} and~\ref{unautrelabel}, with $N=\mathcal{N}/D$). However, Eqs.~\ref{unlabela} and~\ref{unlabel}, and the previous paragraph, show that when $ D$ is increased, the parameter range where the speedup by subdivision tends to the best-scenario value  becomes smaller and smaller. Eventually, this parameter range ceases to exist altogether: this occurs when $R$ becomes of order 1 and below. This sheds light on an interesting trade-off in the degree of subdivision $ D$, between the magnitude of the optimal speedup gained by subdivision and the width of the parameter range over which the actual speedup is close to this optimal value. This effect can be observed qualitatively in Fig.~\ref{OptiSub}A, where the valley crossing time $\tau_m$ of a metapopulation with fixed total size is shown versus the migration-to-mutation rate ratio, $m/(\mu d)$, for different values of $D$: when $D$ is increased, the minimum becomes deeper but less broad.

In addition, Eqs.~\ref{intervs} and~\ref{intervsp} show that when $ D$ is increased, the lower bound of the interval where the speedup by subdivision tends to the best-scenario value decreases, as $D\log D$ for plateaus (Eq.~\ref{intervsp}) and even more rapidly for deep valleys (Eq.~\ref{intervs}). Qualitatively, this is because spreading of the beneficial mutation gets longer when $D$ increases. Conversely, the upper bound of this parameter range is independent of $D$ for deep valleys (Eq.~\ref{intervs}), and grows only logarithmically with $D$ for plateaus (Eq.~\ref{intervsp}). Hence, when $D$ is increased, the center of the interval where the actual speedup is close to the optimal value shifts towards higher migration rates. This effect, which can be observed in Fig.~\ref{OptiSub}A, is studied more precisely in Fig.~\ref{OptiSub}B: at fixed migration rate $m$, the crossing time $\tau_m$ of a metapopulation exhibits a minimum at an intermediate value of $D$. Indeed, the crossing time of the metapopulation first decreases when $D$ is increased because the minimum crossing time then decreases. But beyond a certain value of $D$, the migration rate that yields the highest speedup becomes larger than the fixed migration rate $m$, so $\tau_m$ increases when $D$ is increased further.

Next, we study the dependence on $D$ of the valley crossing time $\tau_\textrm{min}$ minimized over $m$ for each $D$, again for a metapopulation with fixed total size $\mathcal{N}=N D$. For values of $D$ small enough for the interval in Eq.~\ref{interv} to be broad, we expect $\tau_\textrm{min}$ to be close to the optimal scenario value $\tau_{id}/D$. But, as discussed above, as $D$ increases, this interval will become smaller and then vanish. In such a regime, our analytical results are no longer sufficient to predict the dependence of $\tau_\textrm{min}$ on $D$, but our simulations can provide additional insight. Fig.~\ref{OptiSub}C shows that, while $R>100$ (left of the dashed line), $\tau_\textrm{min}$ is close to the best-scenario value. When $D$ is increased beyond this point, $\tau_\textrm{min}$ decreases slower than the best-scenario value. Indeed, the interval in Eq.~\ref{interv} is no longer wide enough for the best-scenario limit to be approached. Note also that when demes become small enough, verifying $N=\mathcal{N}/D\ll1/\delta$ (right of the dotted line in Fig.~\ref{OptiSub}C), mutation `1' becomes effectively neutral in individual demes, as $p_{01}$ tends to $1/N$ (see Eq.~\ref{pij}). For even higher values of $D$, $\tau_\textrm{min}$ is observed to saturate rather than exhibiting a unique minimum. Interestingly, this occurs for $D$ such that the interval in Eq.~\ref{interv} fully vanishes (i.e. when $R$ passes below 1, right of the solid line on Fig.~\ref{OptiSub}C). While we do not have rigorous proof of the generic existence of this saturation, we have explored this point for other parameters, and found similar behavior (data not shown). Importantly, this indicates that there is a whole class of  nearly optimal population structures.

\subsection*{Extension to weakly beneficial intermediates} 
\begin{figure}[h t b]
\centering
\includegraphics[width=0.5\textwidth]{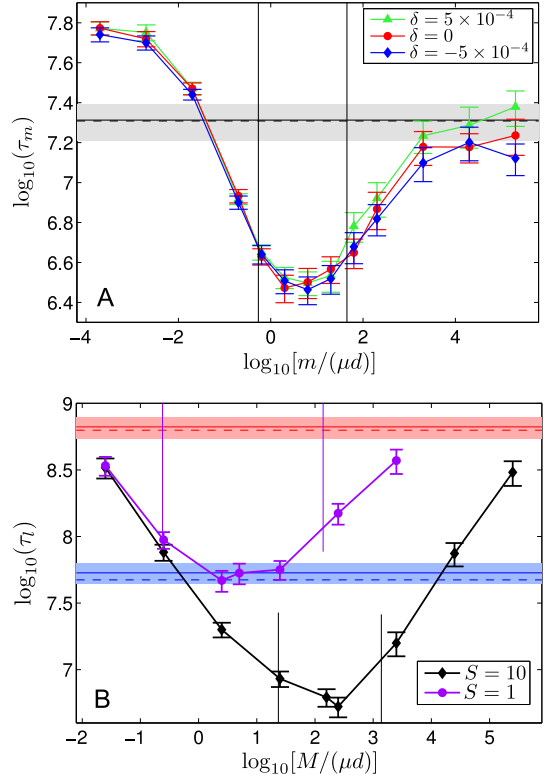}
\caption[Extension to effectively neutral intermediates and to a large population connected to smaller islands.]{\label{Ext}\textbf{Extension to effectively neutral intermediates and to a large population connected to smaller islands.} 
\textbf{A.} Valley crossing time $\tau_m$ of a metapopulation composed of $ D =10$ demes with $K=130$, versus migration-to-mutation rate ratio $m/(\mu d)$, in logarithmic scale, for three values of $\delta$ in the effectively neutral regime. Dots are simulation results, averaged over 100 runs for each value of $m$; error bars represent 95\% CI.  Black vertical lines represent the limits of the interval of $m/(\mu d)$ in Eq.~\ref{interv}. Black horizontal line: plateau crossing time for an isolated deme with $K=130$ for the same parameter values, averaged over 100 runs; shaded regions: 95\% CI. Dashed line: corresponding theoretical prediction from Ref.~\cite{Weissman09}. Note that the plateau crossing time of the non-subdivided population with $K=1300$ is indistinguishable from that of the isolated deme here (as both are in the sequential fixation regime). Parameter values: $d=0.1$, $\mu=5\times10^{-7}$, $s=0.5$; $m$ is varied. \textbf{B.} Valley crossing time $\tau_l$ of a large population with $K=500$ connected to $S$ smaller islands with $K'=50$, versus migration-to-mutation rate ratio $M/(\mu d)$, in logarithmic scale. Dots represent simulation results averaged over 100 runs for each value of $M$, and error bars are 95\% CI. Vertical lines represent the limits of the interval of $M/(\mu d)$ in Eq.~\ref{intervp}. Blue (resp. red) line: valley crossing time for an isolated population with $K=50$ (resp. $K=500$) for the same parameter values, averaged over 100 runs; shaded regions: 95\% CI. Dashed blue (resp. red) lines: corresponding theoretical predictions from Ref.~\cite{Weissman09}. For $S =1$, the observed minimum $\tau_l$ satisfies $\tau_l\approx\tau_{ii}$, where $\tau_{ii}$ is the average valley crossing time of an isolated island. For $S=10$, the observed minimum satisfies $\tau_l\approx\tau_{ii}/10$. Parameter values: $d=0.1$, $\mu=4\times10^{-6}$, $s=0.25$ and $\delta=0.1$; $M$ is varied.}
\end{figure}

Our work has focused on fitness valleys ($\delta>0$), such that mutation `1' is deleterious, and on fitness plateaus ($\delta=0$), such that mutation `1' is neutral. For $|\delta|<\max(\sqrt{\mu s},1/N)$, mutation `1' is effectively neutral, as far as valley crossing is concerned, in a population with $N$ individuals~\cite{Weissman09}. (This condition holds both in the sequential fixation regime and in the tunneling regime.) This implies that our arguments and our results obtained in the case of the fitness plateau also hold for weakly beneficial intermediates. This point is illustrated in Fig.~\ref{Ext}A.

\subsection*{Extension to a population coupled to small island populations} 
Thus far, we focused on demes of equal size for simplicity, but demes of different sizes are relevant in practice. As a step toward more general populations structures, we now consider a population connected by migration to $S$ smaller satellite populations of identical size, assumed to be in the sequential fixation regime. We only allow migration between the large population and each of the smaller islands, and the total migration rate is denoted by $M$. The small island affected by migration is chosen randomly at each migration event. It is straightforward to adapt our work to this case (see Methods, Sec.~\ref{AppC}). We obtain an interval of $M/(\mu d)$ over which the crossing time for the large population is dominated by the crossing time of the champion island. This is corroborated by our simulations (see Fig.~\ref{Ext}B).

\subsection*{Realistic parameter values} 
Let us consider the example of {\it Escherichia coli}, for which the mutation probability per base pair per division is $\mu\approx 8.9\times10^{-11}$~\cite{Wielgoss11}. In order to gain a speedup of crossing by subdivision, we require demes to be in the sequential fixation regime. For plateaus, this condition reads $N<1/\sqrt{\mu s}$. Let us consider deme sizes such that this condition is satisfied.

First, let us choose $N=5\times10^4$, which is within the smallest range of sizes used in current evolution experiments. For instance, it is the number of bacteria transferred at each dilution step for small populations in~\cite{Rozen08}. For this value of $N$, all plateaus with $s<4.5$ are in the sequential fixation regime (from the condition $N<1/\sqrt{\mu s}$). Let us also consider $ D=100$, since 96-well plates are often used in these experiments~\cite{Kerr06,Rozen08}. This yields a total population size of $5\times10^6$ individuals, which is in the tunneling regime for all plateaus with $s>4.5\times 10^{-4}$. For $s=10^{-2}$, isolated demes are in the sequential fixation regime for $0\leq\delta\lesssim2.2\times 10^{-4}$. (Subdivision cannot significantly accelerate crossing for deeper valleys since isolated demes are then in tunneling, but those valleys take longer to cross than shallow ones and are thus probably less often crossed in practice.) The ratio $R$ of the bounds of the interval in Eq.~\ref{interv} satisfies $R>325$ throughout this range of valleys, with $R\approx10^3$ for the plateau and $R>10^4$ for the deepest valleys in the range. Thus,  actual speedups will approach the best-scenario one, and significant speedups will exist in a wide parameter window. Eq.~\ref{lemin} predicts that the highest speedup is obtained for $\delta\approx 3.2\times 10^{-5}$, and Eq.~\ref{lemin2} then yields a speedup factor by subdivision of $\tau_{ns}/\tau_m\approx\tau_{ns}/\tau_c\approx2.9\times 10^2$. (Using instead the full expression of $\tau_c$ obtained from Eq.~\ref{pbest} (see Methods, Sec.~\ref{AppB}) yields $2.7\times 10^2$, i.e. a correction of 7\%.) Moreover, for all valleys with $\delta\leq3.2\times 10^{-5}$, the best-scenario speedup ranges from 18 to $2.7\times 10^2$. Thus, subdivision significantly accelerates crossing for this entire class of valleys. 

It should be noted that the timescales obtained in this example are long compared to experimental ones. For instance, for the plateau, $\tau_c$ corresponds to $1.3\times 10^8$ divisions while $\tau_{ns}$ is $2.4\times 10^9$ divisions. However, $\tau_c$ can become smaller if the number of subpopulations $ D$ is increased, as discussed in our previous section. Besides, we have chosen to focus on standard {\it Escherichia coli} for simplicity. Organisms with a higher mutation rate, e.g. viruses such as HIV, or mutator strains, would have much shorter timescales, but smaller subpopulations would then be required for demes to be in the sequential fixation regime.

Our example thus far  focused on a small but realistic deme size, $N=5\times10^4$. Experimentally more frequent values of $N$ are in the range $5\times10^5$ -- $10^7$~\cite{Kerr06,Rozen08}. Increasing $N$ at fixed $\mu$ decreases the range of $s$ for which demes are in the sequential fixation regime. For a plateau, this condition reads $s<1/(\mu N^2)$. For $N=5\times10^5$, this yields $s<4.5\times 10^{-2}$, and for $N=5\times10^6$, this yields $s<4.5\times 10^{-4}$. Hence, the range of plateaus (and similarly, of valleys) for which subdivision accelerates crossing becomes more restricted when $N$ is increased. Nevertheless, if these increasingly stringent conditions on $s$ are satisfied, significant speedups by subdivision are still expected. Indeed, Eq.~\ref{lemin2} shows that the smallest value of the ratio $\tau_c/\tau_{ns}$ is proportional to $N^2 s$, so if one increases $N$ while decreasing $s$ as $1/N^2$, the maximal speedup by subdivision will remain unchanged.

In this work, we have considered the crossing of one particular valley or plateau corresponding to a specific pair of two mutations. Given the complexity and high dimensionality of actual fitness landscapes, there may be a large number of parallel valleys or plateaus, so that one of these could be crossed quite frequently even though the crossing time for a single valley or plateau remains large. Our work shows that, under specific conditions, subdivision can significantly accelerate crossing for whole classes of valleys and plateaus. Furthermore, in a generic, high-dimensional fitness landscape that contains both valleys and/or plateaus and uphill paths, subdivision can provide an additional effect: it ``shields'' some demes in the metapopulation from adaptation via the uphill paths, leaving them time to explore valley-crossing paths that may be better in the longer term. While this effect is outside the scope of the present paper, it could lead to additional advantages of subdivision in evolution on rugged fitness landscapes.

\subsection*{Conclusion}
Our study of a generic and minimal model of population subdivision with migration demonstrates that subdividing a population into demes connected by migration can significantly accelerate the crossing of fitness plateaus and valleys, without the need for additional ingredients. We have derived quantitative conditions on the various parameters for subdivision to accelerate crossing, and for the resulting speedup to be maximal. In particular, isolated demes have to be in the sequential fixation regime for a significant speedup to occur. This condition is quite strong, but provided that it is met, significant speedups can be obtained in a wide range of migration rates, with the fastest deme driving the crossing of the whole metapopulation in the best scenario. We have derived the interval of migration rates for which this best scenario is reached. In addition, we have shown that increasing the degree of subdivision of a population enables higher speedups to be reached, but that this effect can saturate.

Our quantitative assessment of the conditions under which subdivision significantly speeds up valley or plateau crossing can aid in optimally designing future experiments, enabling one to choose the sizes and the number of demes, as well as the migration rates, such that subdivision can accelerate valley and plateau crossing. 

Further directions include investigating the evolution of a metapopulation with a distribution of deme sizes on a more general rugged landscape, as well as assessing the impact of specific geographic structure. Our work could also be extended to sexual populations, where recombination plays an important role in valley or plateau crossing~\cite{Weissman10}. The interplay between recombination and subdivision, which respectively alleviate and exacerbate clonal interference, would be interesting to study.

\section*{Methods}

\subsection{Simulation methods}
\label{AppSim}

Our simulations are based on a Gillespie algorithm~\cite{Gillespie76,Gillespie77} that we coded in the C language. Here we will describe our algorithm for the case of a metapopulation of $ D$ demes of identical size, which is the primary situation discussed in our work. In our simulations, each deme has a fixed carrying capacity $K$--we discuss this choice further in this section.

\subsubsection{Algorithm}
A number of different events occur in our simulations, each with an independent rate:
\begin{itemize}
 \item Each individual divides at rate $f_g(1-N_i/K)$, where $f_g$ is the fitness associated with the genotype $g\in\{0,1,2\}$ of the individual, and $N_i$ is the current total number of individuals in the deme $i\in[1, D]$ to which the individual belongs. This corresponds to logistic growth. 
\item If a dividing cell has $g<2$, upon division, its offspring (i.e., one of the two individuals resulting from the division) mutates with probability $\mu$, to have genotype $g+1$ instead of $g$. 
\item Each individual dies at rate $d$. Hence, at steady-state, $N_i \approx K(1-d/\bar f_i)$, where $\bar f_i$ is the average fitness of deme $i$. In practice, we choose $d=0.1$, and fitnesses of order one, thus $N_i\approx0.9K$. 
\item Migration occurs at total rate $m \sum_{i=1}^ D N_i$. Two different demes are chosen at random, an individual is chosen at random from each of these two demes, and the two individuals are exchanged. There is no geographic structure in our model, i.e. exchange between any two demes is equally likely. 
\end{itemize}

In practice, the number of individuals with each genotype in each deme is stored, as well as the corresponding division rate. This data fully describes the state of the metapopulation, and allows determination of the rates of all events. For each event in the simulation, the following steps are performed: 
\begin{itemize}
 \item A timestep $dt$ is drawn from an exponential distribution with rate equal to the total rate $R_t$ of events (i.e., the sum of all rates), and time is increased from its previous value, $t$, to $t+dt$. In other words, the next event occurs at time $t+dt$. 
\item The event that occurs at $t+dt$ is chosen randomly, in such a way that the probability of an event with rate $r$ is equal to $r/R_t$: either a cell divides, or a cell dies, or a migration event occurs.
\item The event is performed, and the relevant data is updated. Since we store the number of individuals with each genotype in each deme, only one or two of these numbers need to be updated at each step. In addition, the division rates of the affected deme $i$ must be updated upon division and death because $N_i$ is modified. Note, however, that this represents only three numbers at most (one for each genotype).
\end{itemize}
The advantage of the Gillespie algorithm is that it is exact, and does not involve any artificial discretization of time.

\subsubsection{Working at fixed carrying capacity}
\label{AppSimK}

In our simulations, demes have a fixed carrying capacity, and the number of individuals per deme fluctuates weakly around its equilibrium value. This approach, also used in e.g.~\cite{Greulich12}, has the advantage of realism. Alternatively, we could impose a constant number of individuals per deme. \\(i) First, we could choose a dividing individual in the whole metapopulation with probability proportional to its fitness, and simultaneously suppress another individual, chosen at random in the same deme. However, in this case, individuals in demes of higher fitness would exhibit shorter lifespans, which is not realistic and may introduce a bias. \\(ii) A second possibility would be to choose a dividing individual (according to fitness) in each of the demes, and to simultaneously suppress another individual, chosen at random, in each deme. However, in this case, unless migration events are far less frequent than these collective division-death events (i.e., these $ D$ division-death events), the time interval between them becomes artificially discretized. This introduces biases unless the total migration rate $m D N$ is much smaller than $Nd$, i.e. unless $m\ll d/ D$. 

Consequently, while imposing a constant number of individuals is a good simulation approach for a non-subdivided population (see e.g.~\cite{Weissman09}), it tends to introduce biases in the study of metapopulations. While we chose to perform simulations with fixed carrying capacities in order to avoid any of these biases, we checked that, for small enough migration rates, our results are completely consistent with simulation scheme (ii) described above. This consistency check also demonstrates that it is legitimate to compare our simulation results obtained with fixed carrying capacities to our analytical work carried out with constant population size per deme.

\subsection{Crossing time of the champion deme}
\label{AppB}

In this section, we give more details on the calculation of the average valley or plateau crossing time $\tau_c$ by the champion deme amongst $ D$ independent ones. We show in the Results section that, in the best scenario, the crossing time of the whole metapopulation is determined by this time.

$\tau_c$ is the average shortest crossing time of $ D $ independent demes. This minimum crossing time, which we denote by $t_c$, is also called the smallest (or first) order statistic of the deme crossing time amongst a sample of size $ D $~\cite{Bolch06}. 

Let us denote by $p(t)$ the probability density function of valley or plateau crossing time for a single deme, and let us introduce $\mathcal{P}(t)=\int_t^\infty p(t') dt'$ (it satisfies $\mathcal{P}(t)=1-\mathcal{C}(t)$ where $\mathcal{C}(t)$ is the cumulative distribution function of valley or plateau crossing by a single deme). The probability that $t_c$ is larger than $t$ is equal to the probability that the crossing times of each of the $ D$ independent demes are all larger than $t$: $P(t_c\geq t)=\left[\mathcal{P}(t)\right]^ D$. By differentiating this expression, one obtains the probability density function $p_c(t_c)$ of the crossing time $t_c$ by the champion deme~(see e.g.~\cite{Bolch06}):
\begin{equation}
p_c(t_c)= D \left[\mathcal{P}(t_c)\right]^{ D-1} p(t_c)\,. \label{pb}
\end{equation}

We now express $p(t)$ explicitly. Since demes are assumed to be in the sequential fixation regime, valley or plateau crossing involves two successive steps. The first step, fixation of a `1'-mutant, occurs with rate $r_{01}$, and the second step, fixation of a `2'-mutant, occurs with rate $r_{12}$ (see the Results section for expressions of these rates). The total crossing time is thus a sum of two independent exponential random variables, with probability density function given by a two-parameter hypoexponential distribution~\cite{Bolch06}:
\begin{equation}
p(t)=\frac{r_{01}r_{12}}{r_{12}-r_{01}}\left(e^{-r_{01}t}-e^{-r_{12}t}\right)\,. \label{hypo}
\end{equation}
Combining Eqs.~\ref{pb} and~\ref{hypo}, we obtain
\begin{equation}
p_c(t_c)= D\left(\frac{r_{12}e^{-r_{01}t_c}-r_{01}e^{-r_{12}t_c}}{r_{12}-r_{01}}\right)^{ D-1}p(t_c)\,, \label{pbest}
\end{equation}
with $p(t_c)$ given by Eq.~\ref{hypo}. $\tau_c$ can then be determined for any value of the parameters by computing the average value of $t_c$ over this distribution.

Since mutation `1' is deleterious or neutral while mutation `2' is beneficial, the first step of valley crossing is much longer than the second one over a broad range of parameter values. In this case, we can approximate $p(t)$ with a simple exponential distribution,
\begin{equation}
p(t)=r_{01}e^{-r_{01}t}\,.
\end{equation}
Eq.~\ref{pb} then yields
\begin{equation}
p_c(t_c)= D r_{01}e^{- D r_{01}t_c}\,,
\end{equation}
i.e. $t_c$ is distributed exponentially with rate $ D r_{01}$. In this case, we simply have $\tau_c\approx1/( D r_{01})$, which can be written as $\tau_c\approx\tau_{id}/ D $, where $\tau_{id}\approx 1/r_{01}$ is the average crossing time for an isolated deme. Hence, in this case, on which our analytical discussion focus, the champion deme crosses the valley $ D $ times faster on average than an isolated deme.

For this approximation to be valid, the second step of valley crossing must be negligible even for the champion deme, i.e., $ D p_{01}\ll p_{12}$. For very large $ D$, the value of $\tau_c$ will not be as small as $1/( D r_{01})$, since the second step will no longer be negligible (see~\cite{Weissman10} for a discussion of similar issues). The crossover to this regime can be determined by computing the average of the distribution in Eq.~\ref{pbest} and comparing it to $1/( D r_{01})$.

\subsection{Number of migration events for extinction or spreading in a metapopulation}
\label{AppA}

In our Results section, we have derived an interval of the ratio of migration rate to mutation rate over which subdivision most reduces valley or plateau crossing time (see Eq.~\ref{interv}). The upper bound involves $n_e$, the average number of migration events required for the `1'-mutants to be wiped out by migration, starting from a state where one deme has fixed genotype `1', while all other demes have genotype `0'. Similarly, the lower bound involves $n_s$, the average number of migration events required for the `2'-mutants to spread by migration to the whole metapopulation, starting from a state where one deme has fixed genotype `2', while all other demes have genotype `0'. In our Results section, we have provided intuitive derivations of the simple expressions of $n_s$ and $n_e$, valid for $Ns\gg 1$ and $s\ll 1$, $N\delta\gg 1$ and $\delta\ll1$ (see Eq.~\ref{intervs}). However, it is important to derive more general expressions, especially since subdivision generically most accelerates valley crossing in the intermediate regime where $N\delta\approx1$ (see Results, Eq.~\ref{lemin}). 

Here, we derive general analytical expressions for $n_e$ and $n_s$, both for fitness plateaus and for fitness valleys. These more general expressions are those used for numerical calculations of the bounds in our examples. Throughout this section, we consider a metapopulation of $ D$ demes composed of $N$ individuals each, and we assume that individual demes are in the sequential fixation regime (see Results).

\subsubsection{A finite Markov chain}

In order to determine $n_s$ and $n_e$, we study the evolution of the number $i\in [0, D ]$ of demes that have fixed the mutant genotype (`1' for the calculation of $n_e$; `2' for that of $n_s$), while other demes have genotype `0'. Given that the value of $i$ just before a migration step fully determines the probabilities of the outcomes of this migration step, and given that $i=0$ and $i= D$ are absorbing states, the number $i$ evolves according to a finite Markov chain, each step being a migration event. We next express the transition matrix of this Markov chain. 

The only migration events that can affect $i$ are those that exchange individuals from two demes with different genotypes. Let us call these migration events ``relevant''. The probability $p^r_i$ of a migration event being relevant corresponds to the probability that this migration affects one of the $i$ mutant populations and one of the $ D-i$ `0' populations: $p^r_i=2i( D -i)/[ D ( D -1)]$. We only focus on the final outcome of a migration event, after fixation or extinction of each of the two migrants' lineages has occurred. Let $p$ denote the probability that the mutant migrant fixes in the `0' deme, and $p'$ the probability that the `0' migrant fixes in the mutant deme. As a result of one such relevant migration event:
\begin{itemize}
\item $i$ increases by one with probability $p(1-p')$, if the migrant mutant fixes in the `0' deme while `0' migrant does not fix in the mutant deme.
\item $i$ decreases by one with probability $p'(1-p)$, in the opposite case.
\item Otherwise, $i$ does not change. This happens either if both migrants fix (with probability $pp'$) or if no migrant fixes (with probability $(1-p)(1-p')$).
\end{itemize}
These probabilities, multiplied by the probability $p^r_i$ that a migration event is relevant, yield the transition matrix of our finite Markov chain, which is tri-diagonal (or continuant) since each migration step can either leave $i$ constant, or increase or decrease it by one:
\begin{align}
P_{i\rightarrow i+1}&=\frac{2i( D -i)}{ D ( D -1)}p(1-p')\,,\label{Pincr}\\
P_{i\rightarrow i-1}&=\frac{2i( D -i)}{ D ( D -1)}p'(1-p)\,,\label{Pdecr}\\
P_{i\rightarrow i}&=1-P_{i\rightarrow i+1}-P_{i\rightarrow i-1}\,,
\end{align}
for $i\in[1, D-1]$, and $P_{0\rightarrow 0}=P_{ D\rightarrow  D}=1$. We have denoted by $P_{j\rightarrow k}$ the probability that $i$ varies from $j$ to $k$ as the final outcome of one migration event.  

Here, we do not account for independent mutations arising and fixing in other demes during the process of spreading (or extinction) of the mutant's lineage in the metapopulation. Indeed, our aim is to compare the timescales of migration and mutation processes, so we treat them separately. Note that, in practice, this hypothesis is reasonable if mutations that fix are sufficiently rarer than migration events. We also consider that the time between two successive migration events is large enough for fixation to occur in the demes affected by migration before the next migration event occurs, which is true in the low-migration rate regime that we study in our work ($2m\ll d$, where $2m$ is the migration rate per individual, while $d$ is the death and division rate per individual).

$n_s$ and $n_e$ can be directly expressed as the average number of steps of the Markov chain necessary to go from the initial state $i=1$ to absorption in a particular absorbing state, either $i= D$ or $i=0$. Let us present general expressions of these average numbers of steps, before using them to obtain explicit expressions of $n_s$ and $n_e$.

\subsubsection{Some results regarding finite Markov chains with tri-diagonal probability matrices}
\label{Mark}

We are interested in the average number of steps $\nu_a$ until the system reaches each of the absorbing states $a\in\{0, D\}$, starting from the state $i=1$:
\begin{equation}
\nu_a=\sum_{j=1}^{ D-1}s_{j,a}\,,
\end{equation}
where $s_{j,a}$ is the average number of steps that the system spends in the state $i=j$ before absorption, given that it starts in the state $i=1$ and finally absorbs in state $i=a$. It can be expressed as~\cite{Ewens79}
\begin{equation}
s_{j,a}=\frac{\pi_{j,a}}{\pi_{1,a}}s_{j}\,,
\end{equation}
where $s_{j}$ is the average number of steps the system spends in state $i=j$ before absorption in either of the two absorbing states, given that it started in state $i=1$, and $\pi_{j,a}$ is the probability that the system finally absorbs in state $i=a$ if it starts in state $i=j$. 

Using the explicit expressions given in~\cite{Ewens79} for $s_{j}$ and $\pi_{j,a}$ in the case of a tri-diagonal probability matrix, we obtain:
\begin{align}
\nu_0&=\sum_{j=1}^{ D-1}\frac{\left(\sum_{k=j}^{ D-1}\rho_k\right)^2}{\left(\sum_{k=0}^{ D-1}\rho_k\right)\left(\sum_{k=1}^{ D-1}\rho_k\right)\rho_jP_{j\rightarrow j+1}}\,,
\label{n10}\\
\nu_ D&=\sum_{j=1}^{ D-1}\frac{\left(\sum_{k=0}^{j-1}\rho_k\right)\left(\sum_{k=j}^{ D-1}\rho_k\right)}{\left(\sum_{k=0}^{ D-1}\rho_k\right)\rho_jP_{j\rightarrow j+1}}\,,
\label{n1N}
\end{align}
where we have introduced 
\begin{equation}
\rho_k=\prod_{i=1}^{k}\frac{P_{i\rightarrow i-1}}{P_{i\rightarrow i+1}}\,. 
\label{rho}
\end{equation}

\subsubsection{Explicit expression of $n_e$}

$n_e$, in fact, corresponds to $\nu_0$, where $p$ is the probability that a `1'-mutant fixes in a deme of `0' individuals (i.e. $p=p_{01}$) and $p'$ is the probability that a `0'-individual fixes in a deme of `1'-mutants (i.e. $p'=p_{10}$). Hence, it can be expressed explicitly from Eqs.~\ref{n10}, \ref{Pincr}, and \ref{Pdecr}. Since the expressions of $p$ and $p'$ depend whether mutation `1' is neutral or deleterious, we obtain different expressions for the fitness plateau and for the fitness valley. 

\paragraph*{Fitness plateau.} For a fitness plateau (i.e. a neutral intermediate `1'), $p=p'=1/N$, where $N$ is the number of individuals per deme. Hence,
\begin{equation}
P_{i\rightarrow i+1}=P_{i\rightarrow i-1}=\frac{2i( D -i)(N-1)}{ D( D-1)N^2}\,,
\end{equation}
which implies that $\rho_k=1$ for all $k$ (see Eq.~\ref{rho}). Thus, Eq.~\ref{n10} yields
\begin{equation}
n_e=\frac{N^2  D }{2(N-1)}\sum_{j=2}^{ D }\frac{1}{j}\approx \frac{N}{2} D\log D\,,
\label{n1n0}
\end{equation}
where the last expression holds for $N\gg 1$ and $ D\gg 1$.

\paragraph*{Fitness valley.} Eqs.~\ref{rho} and~\ref{Pincr},~\ref{Pdecr} yield $\rho_k=r^k$, with 
\begin{equation}
r=\frac{(1-p)p'}{(1-p')p}\,, \label{r}
\end{equation}
and Eq.~\ref{n10} gives:
\begin{equation}
n_{e}=\frac{ D ( D -1)}{2(r-r^{ D })(1-r^{ D })(1-p')p}\sum_{j=1}^{ D -1}\frac{(r^j-r^{ D })^2}{r^j j( D -j)}\,.
\label{n1d0}
\end{equation}

In these expressions, $p= p_{01}$ is the probability of fixation of a deleterious `1'-mutant, with fitness $1-\delta$, in a deme where all other individuals have genotype `0' and fitness $1$. It can be obtained from Eq.~\ref{pij}, as well as the probability $p'= p_{10}$ of the opposite process.

\subsubsection{Explicit expression of $n_s$}

$n_s$ corresponds to $\nu_ D$, where $p= p_{02}$ is the probability that a `2'-mutant (with fitness $1+s$) fixes in a deme of `0' individuals (with fitness 1), and $p'= p_{20}$ is the probability that a `0'-individual fixes in a deme of `2'-mutants. Hence, it can be expressed explicitly from Eqs.~\ref{n1N}, \ref{Pincr}, and \ref{Pdecr}, using Eq.~\ref{pij} to express the fixation probabilities. For $n_s$, there is no difference between the valley and the plateau, since genotype `1' is not involved. 

As above, Eqs.~\ref{Pincr}, \ref{Pdecr} and~\ref{rho} yield $\rho_k=r^k$, with $r$ defined in Eq.~\ref{r}. Thus, Eq.~\ref{n1N} gives
\begin{equation}
n_s=\frac{ D ( D -1)}{2(1-r)(1-r^{ D })(1-p')p}\sum_{j=1}^{ D -1}\frac{(1-r^j)(1-r^{ D -j})}{j( D -j)}\,.
\label{n1bN}
\end{equation}

\subsubsection{Simplified expressions for deep valleys and for plateaus}

In our Results section, we have shown that the benefit of subdivision is highest when $m/(\mu d)$ is situated between a lower bound,
\begin{equation}
L=n_s p_{01}\,,\label{L}
\end{equation}
and an upper bound,
\begin{equation}
U=\frac{n_e p_{12}}{ D }\,,\label{U}
\end{equation}
(see Eq.~\ref{interv}), where $p_{12}$ denotes the probability of fixation of a single mutant with genotype `2' in a background of `1'-mutants. Here we present simplified expressions for $n_s$ and $n_e$, and hence of $L$ and $U$, in particular parameter regimes. 

Throughout this section, we focus on the regime where $Ns\gg1$ but $s\ll1$, such that mutation `2' is substantially, but not overwhelmingly, beneficial~\cite{Weissman09}. We then have $p_{20}\approx se^{-Ns}\ll1$ and $p_{02}\approx s$ (see Eq.~\ref{pij}). To leading (i.e. zeroth) order in $p_{20}$, we obtain from Eq.~\ref{n1bN} that
\begin{equation}
n_s=\frac{ D -1}{s}\sum_{j=1}^{ D -1}\frac{1}{j}\approx \frac{ D\log D}{s}\,,
\label{nsSimpl}
\end{equation}
where the last expression holds for $ D\gg 1$. This expression of $n_s$ is identical to Eq.~\ref{ns}, which was demonstrated more intuitively in the Results section by directly assuming $Ns\gg1$ and $s\ll1$.

We now consider the case of a plateau ($\delta=0$) and the case of a valley such that $N \delta\gg1$ but $\delta\ll1$. We demonstrate that the latter case is consistent with the simplified derivations in our Results section. 

\paragraph*{Fitness plateau.}

For a fitness plateau, combining Eqs.~\ref{n1n0} and~\ref{U}, the upper bound $U$ reads
\begin{equation}
U\approx\frac{N s }{2}\log  D\,,
\label{UP}
\end{equation}
where we have used $p_{12}=p_{02}\approx s$, since mutation `1' is neutral, and assumed $N\gg 1$ and $ D  \gg 1$.

Additionally, Eqs.~\ref{nsSimpl} and~\ref{L} can be combined to write the lower bound $L$ as
\begin{equation}
L\approx\frac{ D }{Ns}\log  D\,.
\label{LP}
\end{equation}
again to lowest order in $p_{20}$. Here, we have used $p_{01}=1/N$, since in the case of the plateau, mutation `1' is neutral. This expression too holds for $N\gg 1$ and $ D  \gg 1$. 

Combining Eqs.~\ref{UP} and~\ref{LP} yields Eq.~\ref{intervsp}.

\paragraph*{Fitness valley.}

Next we focus on valleys such that $\delta\ll1$ but $N\delta\gg1$. (Note that, in the opposite limit $N\delta\ll1$, mutation `1' is effectively neutral, and the above discussion regarding the fitness plateau applies.) Then, $p_{01}\approx \delta e^{-N\delta}\ll1$ (see Eq.~\ref{pij}). To lowest (i.e. zeroth) order in $p_{01}$, Eq.~\ref{n1d0} becomes
\begin{equation}
n_e\approx \frac{ D}{2p_{10}}\approx \frac{ D}{2\delta}\,,
\end{equation}
where we have used the approximation $p_{10}\approx\delta$, which holds for $\delta\ll1$ and $N\delta\gg1$. This expression of $n_e$ coincides with Eq.~\ref{nd}, which is obtained in the Results section through a more intuitive argument that directly assumes $\delta\ll1$ and $N\delta\gg1$. Hence, from Eq.~\ref{U}, the upper bound $U$ is
\begin{equation}
U\approx\frac{p_{12}}{2\delta}\approx\frac{1}{2}\left(1+\frac{s}{\delta}\right)\,, \label{UV}
\end{equation}
where we used the conditions $\delta\ll1$, $N\delta\gg1$, $s\ll1$ and $Ns\gg1$ to simplify the expression of $p_{12}$.

Meanwhile, from Eq.~\ref{L} and~\ref{nsSimpl}, the lower bound $L$ takes the form
\begin{equation}
L=\frac{p_{01}}{p_{02}} D \log D\approx\frac{\delta e^{-N\delta}}{s} D \log D\,, \label{LV}
\end{equation}
where, again, we used the conditions $\delta\ll1$, $N\delta\gg1$, $s\ll1$ and $Ns\gg1$ to simplify the expressions of $p_{01}$ and $p_{02}$.

Combining Eqs.~\ref{LV} and~\ref{UV} yields Eq.~\ref{intervs}.

\subsection{A population connected by migration to smaller population islands}
\label{AppC}

Let us consider a population of $\mathcal{N}$ individuals connected by migration to $S$ smaller population islands with $N<\mathcal{N}$ individuals each. These islands of identical size are assumed to be in the sequential fixation regime. For the sake of simplicity, we consider that migration only occurs between the large population and the islands: a migration step is a random exchange of two individuals between the large population and one of the islands (chosen at random at each migration event), and the total migration rate is denoted by $M$. Here, we focus on the valley or plateau crossing time of the large population. We demonstrate that the evolution of a large population can be driven by that of satellite islands.

In the optimal case, the crossing time of the large population is determined by that of the champion island, i.e., that which crosses the fitness valley or plateau fastest. We now determine the conditions under which this optimum is achieved, focusing on migration rates much smaller than division/death rates, $M\ll \min(d N  S, d \mathcal{N})$, such that fixation or extinction of a mutant lineage in either the large population or an island is not significantly perturbed by migration. Again, migration should be rare enough for islands to remain effectively shielded from migration events while they have fixed the intermediate mutation, until the final beneficial mutation arises. Second, migration should also be frequent enough for the spreading time of the final beneficial mutation from the champion island to the large population to be negligible with respect to the crossing time of the champion island. These two criteria again provide upper and lower bounds on $M/(\mu d)$. 

The average time $\tau_{12}=1/(N\mu d p_{12})$ (with $p_{12}$ from Eq.~\ref{pij}) required for an island of `1'-mutants to fix the beneficial mutation `2' must be smaller than the average time, $t_{e}$, for an island of `1'-mutants to be wiped out by migration from the large population, which still exhibits genotype `0'. The rate of migration events between the island of `1'-mutants and the large population is $M/S$. Hence, $t_{e}=S /(M p_{10})$, where $p_{10}$ is the probability of fixation of the lineage of a single migrant with genotype `0' in an island where all other individuals are `1'-mutants: for valleys, it is given by Eq.~\ref{pij}, while for plateaus, it is equal to $1/N$. The first condition, $\tau_{12}< t_e$, thus yields
\begin{equation}
\frac{M}{\mu d}<\frac{N S p_{12}}{p_{10} }\,.
\label{TSp}
\end{equation} 

The second condition is that the average spreading time, $t_s$, for the final beneficial mutation to fix in a large population after it has fixed in the champion island, must be smaller than the average valley or plateau crossing time, $\tau_c$, of the champion island. Similar to $t_e$ previously, we obtain $t_s=S/(Mp_{02}^l)$, where $p_{02}^l=(1-e^{-s})/(1-e^{-\mathcal{N}s})$ is the probability of fixation of a migrant with genotype `2' in the large population, which is assumed to exhibit genotype `0' before migration (see Eq.~\ref{pij}). $\tau_c$ is the average of the minimum crossing time among $S$ independent islands. We again focus, for simplicity, on the limit where the first step of valley or plateau crossing, which occurs at rate $r_{01}$, is much longer than the second. Then, we simply have $\tau_c\approx\tau_{ii}/S $ (see Results). In this expression, $\tau_{ii}\approx r_{01}^{-1}=1/(N\mu d p_{01})$ (with $p_{01}$ obtained from Eq.~\ref{pij}) is the average crossing time for an isolated island. Hence, the champion island crosses the valley $S$ times faster on average than a single isolated island. The second condition, $t_s<\tau_c$, finally yields
\begin{equation}
\frac{S^2Np_{01}}{p_{02}^l}<\frac{M}{\mu d}\,.
\label{TIp}
\end{equation}

Together, Eqs.~\ref{TSp} and~\ref{TIp} yield the interval of $M/\mu d$ over which we expect subdivision to maximally accelerate crossing:
\begin{equation}
\frac{S^2Np_{01}}{p_{02}^l}\ll\frac{M}{\mu d}\ll\frac{N S p_{12}}{p_{10} }\,.
\label{intervp}
\end{equation}
In this range, we expect the valley or plateau crossing time $\tau_l$ of the large population to be dominated by the crossing time of the champion island, so that $\tau_l/\tau_{ii}\approx1/S$. This prediction is confirmed by our simulations (see Fig.~\ref{Ext}B).

\section*{Acknowledgments}
We thank Ned S. Wingreen for stimulating discussions. AFB acknowledges the support of the Human Frontier Science Program. DJS was supported by National Institutes of Health Grant K25 GM098875. This work was also supported in part by National Science Foundation Grant PHY-0957573 and National Institutes of Health Grant R01 GM082938.


\section*{Figures}

\end{document}